\documentclass{emulateapj}
\usepackage[english]{babel}
\usepackage[autostyle, english = american]{csquotes}
\usepackage{fancyhdr}
\usepackage{booktabs,longtable}
\usepackage{hyperref}
\usepackage{upgreek}
\usepackage[mediumspace,mediumqspace,squaren]{SIunits}
\bibpunct{(}{)}{;}{a}{}{,} 
\newcommand{\HI}{{H{\,\footnotesize I~}}}
\newcommand{\kms}{km s$^{-1}$~}

\newcommand{\Cq}{$z=0.376$}

\begin{document}
\title{Highest redshift image of neutral hydrogen in emission: A CHILES detection of a starbursting galaxy at \Cq}
\author{Ximena Fern\'{a}ndez\altaffilmark{1,2,*},  Hansung B. Gim\altaffilmark{3}, J.H. van Gorkom\altaffilmark{2}, Min S. Yun\altaffilmark{3}, Emmanuel Momjian\altaffilmark{4}, Attila Popping\altaffilmark{5,6},  Laura Chomiuk\altaffilmark{7},  Kelley M. Hess\altaffilmark{8,9}, Lucas Hunt\altaffilmark{10}, Kathryn Kreckel\altaffilmark{11}, Danielle Lucero\altaffilmark{8}, Natasha Maddox\altaffilmark{9}, Tom Oosterloo\altaffilmark{9,8}, D.J. Pisano\altaffilmark{10}, M.A.W. Verheijen\altaffilmark{8}, Christopher A. Hales\altaffilmark{4,+}, Aeree Chung\altaffilmark{12}, Richard Dodson\altaffilmark{5}, Kumar Golap\altaffilmark{4}, Julia Gross\altaffilmark{2}, Patricia Henning\altaffilmark{13}, John Hibbard\altaffilmark{14}, Yara L. Jaff\'{e}\altaffilmark{15}, Jennifer Donovan Meyer\altaffilmark{14}, Martin Meyer\altaffilmark{5}, Monica Sanchez-Barrantes\altaffilmark{13}, David Schiminovich\altaffilmark{2}, Andreas Wicenec\altaffilmark{5}, Eric Wilcots\altaffilmark{16},  Matthew Bershady\altaffilmark{16}, Nick Scoville\altaffilmark{17}, Jay Strader\altaffilmark{7}, Evangelia Tremou\altaffilmark{7}, Ricardo Salinas\altaffilmark{7}, Ricardo Ch{\'a}vez\altaffilmark{18}}
\altaffiltext{1}{Department of Physics and Astronomy, Rutgers, The State University of New Jersey, Piscataway, NJ 08854-8019, USA}
\altaffiltext{2}{Department of Astronomy, Columbia University, New York, NY 10027, USA}
\altaffiltext{3}{Department of Astronomy, University of Massachusetts, Amherst, MA 01003, USA}
\altaffiltext{4}{National Radio Astronomy Observatory, PO Box 0, Socorro, NM 87801, USA}
\altaffiltext{5}{International Centre for Radio Astronomy Research, The University of Western Australia, Crawley, WA 6009, Australia}
\altaffiltext{6}{Australian Research Council, Centre of Excellence for All-sky Astrophysics (CAASTRO)}
\altaffiltext{7}{Department of Physics and Astronomy, Michigan State University, East Lansing, MI 48824, USA}
\altaffiltext{8}{Kapteyn Astronomical Institute, University of Groningen, Groningen, The Netherlands}
\altaffiltext{9}{Netherlands Institute for Radio Astronomy (ASTRON), Dwingeloo, The Netherlands}
\altaffiltext{10}{Department of Physics and Astronomy, West Virginia University, P.O. Box 6315, Morgantown, WV 26506, USA}
\altaffiltext{11}{Max Planck Institute for Astronomy, Königstuhl 17, D-69117 Heidelberg, Germany}

\altaffiltext{12}{Department of Astronomy, Yonsei University, Seoul 120-749, Republic of Korea}
\altaffiltext{13}{Department of Physics and Astronomy, University of New Mexico, Albuquerque, NM 87131, USA}
\altaffiltext{14}{National Radio Astronomy Observatory, Charlottesville, VA 22903 USA}
\altaffiltext{15}{European Southern Observatory, Alonso de Cordova 3107, Vitacura, Santiago, Chile}
\altaffiltext{16}{Department of Astronomy, University of Wisconsin-Madison, Madison, WI 53706, USA}
\altaffiltext{17}{Department of Astronomy, California Institute of Technology, Pasadena, CA 91125, USA}
\altaffiltext{18}{Instituto Nacional de Astrof{\'i}sica {\'O}ptica y Electr{\'o}nica, AP 51 y 216, 72000, Puebla, M{\'e}xico}
\altaffiltext{*}{NSF Astronomy and Astrophysics Postdoctoral Fellow}
\altaffiltext{+}{Jansky Fellow of the National Radio Astronomy Observatory}
\begin{abstract}
Our current understanding of galaxy evolution still has many uncertainties associated with the details of accretion, processing, and removal of gas across cosmic time. The next generation of radio telescopes will image the neutral hydrogen (H{\footnotesize I}) in galaxies over large volumes at high redshifts, which will provide key insights into these processes. We are conducting the COSMOS \HI Large Extragalactic Survey (CHILES) with the Karl G. Jansky  Very Large Array, which is the first survey to simultaneously observe \HI from $z=0$ to $z\sim0.5$. Here, we report the highest redshift \HI 21-cm detection in emission to date of the luminous infrared galaxy (LIRG) COSMOS J100054.83+023126.2 at z=0.376 with the first 178 hours of CHILES data. The total \HI mass is $(2.9\pm1.0)\times10^{10}~M_\odot$, and the spatial distribution is asymmetric and extends beyond the galaxy. While optically the galaxy looks undisturbed, the \HI distribution suggests an interaction with candidate a candidate companion. In addition, we present follow-up Large Millimeter Telescope CO observations that show it is rich in molecular hydrogen, with a range of possible masses of $(1.8-9.9)\times10^{10}~M_\odot$. This is the first study of the \HI and CO in emission for a single galaxy beyond $z\sim0.2$. 
\end{abstract}

\section{Introduction}
Galaxy evolution studies have been hampered by the lack of neutral hydrogen (\HI) images across cosmic time. The \HI 21-cm line has been used extensively to study nearby galaxies since it is the raw fuel for star formation, probes internal properties, and serves as an excellent environmental tracer \citep[e.g.,][]{Haynes84,THINGS,Chung09}. Due to technical limitations, not much is known about \HI in distant galaxies but this will soon change with the next generation of radio telescopes.
\HI imaging at $z>0.1$ will help constrain the evolution of the interstellar medium (ISM), and possibly explain observed phenomena at high redshift, such as the decline in the star formation rate (SFR) since $z=2$ \citep{Hopkins06}.

Over the past 15 years, several studies have started to uncover the \HI content beyond $z\sim0.1$.  Targeted surveys have observed \HI for a limited number of galaxies at higher redshift \citep[e.g.,][]{Zwaan01, Catinella15}. 
Long integration times are necessary to detect \HI emission from distant galaxies since the signal is very weak. As a consequence, many studies use indirect methods such as stacking and intensity mapping to attain a statistical measure of how much \HI there is in the interval $z\sim0.1-0.8$ \citep[e.g.,][]{Lah07, Chang10, Masui13, Delhaize13, Gereb13}. 

New technology now allows telescopes to carry out \HI observations with large instantaneous frequency coverage. The first survey to do this is the Blind Ultra Deep \HI Environmental Survey (BUDHIES), which detected \HI in over 150 galaxies in and around two clusters at $0.16<z<0.22$ \citep{Verheijen07}. The recently upgraded Karl G. Jansky Very Large Array (VLA)\footnote{The National Radio Astronomy Observatory is a facility of the National Science Foundation operated under cooperative agreement by Associated Universities, Inc.} can now observe the interval $0<z<0.5$ in one setting. We did a pilot study during commissioning that covered the interval $0<z<0.193$ \citep{Fernandez13}. We reached the theoretical noise over the entire frequency range, demonstrating the feasibility of an \HI deep field.  We are now conducting the COSMOS \HI Large Extragalactic Survey (CHILES) with the VLA, a 1002 hr survey where we expect to directly image the \HI distribution and kinematics of at least 300 galaxies in the COSMOS field from $z=0$ to $z\sim0.5$, with at least 200 of these at $z>0.2$.  These estimates are done by comparing our $5\sigma$ sensitivity curve (assuming a 150 \kms width) to the \HI masses predicted for spectroscopically confirmed galaxies in the observed volume using the scaling relation from  \citet{Catinella12}. 

Here, we report the \HI detection of the Luminous Infrared Galaxy (LIRG) COSMOS J100054.83+023126.2 (henceforth J100054) at $z=0.376$ with the first 178 hr of CHILES data.  In addition, we include  CO observations from the Large Millimeter Telescope Alfonso Serrano (LMT), and follow-up optical spectroscopic data that confirm the redshift of the detection. This study not only represents the highest redshift \HI detection in emission,  but is the first time we can study the cold gas content (both molecular and atomic hydrogen) of a galaxy at $z>0.2$. 

This letter adopts $H_{\rm o}=70$ km s$^{-1}$ Mpc$^{-1}$, $\Omega_M=0.3$, and  $\Omega_\Lambda=0.7$, and a \citet{Kroupa01} initial mass function.

\section{Observations and Results}
\subsection{Cold gas properties}
\subsubsection{CHILES: HI detection at z=0.376}
We are observing one pointing in the COSMOS field with the B configuration of the VLA, which corresponds to a spatial resolution of $\sim25$ kpc at $z=0.376$. The first 178 hr (Phase I) were observed in 2013 (Momjian et al, in prep).

For this preliminary study, we imaged the interval $0.36<z<0.39$ ($1025-1045$ MHz) which includes a wall with over 250 galaxies with spectroscopically known redshifts, and 60 of these have predicted  $M_{\rm HI}> 10^{10}~M_\odot$.  The cube was made with  robustness parameter of 0.8 in CASA, has a spatial scale of $2048\times2048$ of 2$\arcsecond$ pixels, and a total of 320 channels of 62.5 kHz ($18$  km s$^{-1}$ at $z=0.376$). We first Hanning smooth the data to increase the S/N and then iteratively subtract the continuum in the image plane, leading to a typical noise per channel of 75 $\micro$Jy beam$^{-1}$. We search for \HI by eye around a subset of the 60 predicted gas-rich galaxies. We are still in the process of searching for \HI around other galaxies directly and via stacking.

\begin{figure*}
\begin{center}
\includegraphics[trim=18mm 5mm 10mm 0mm,clip,scale=1.0]{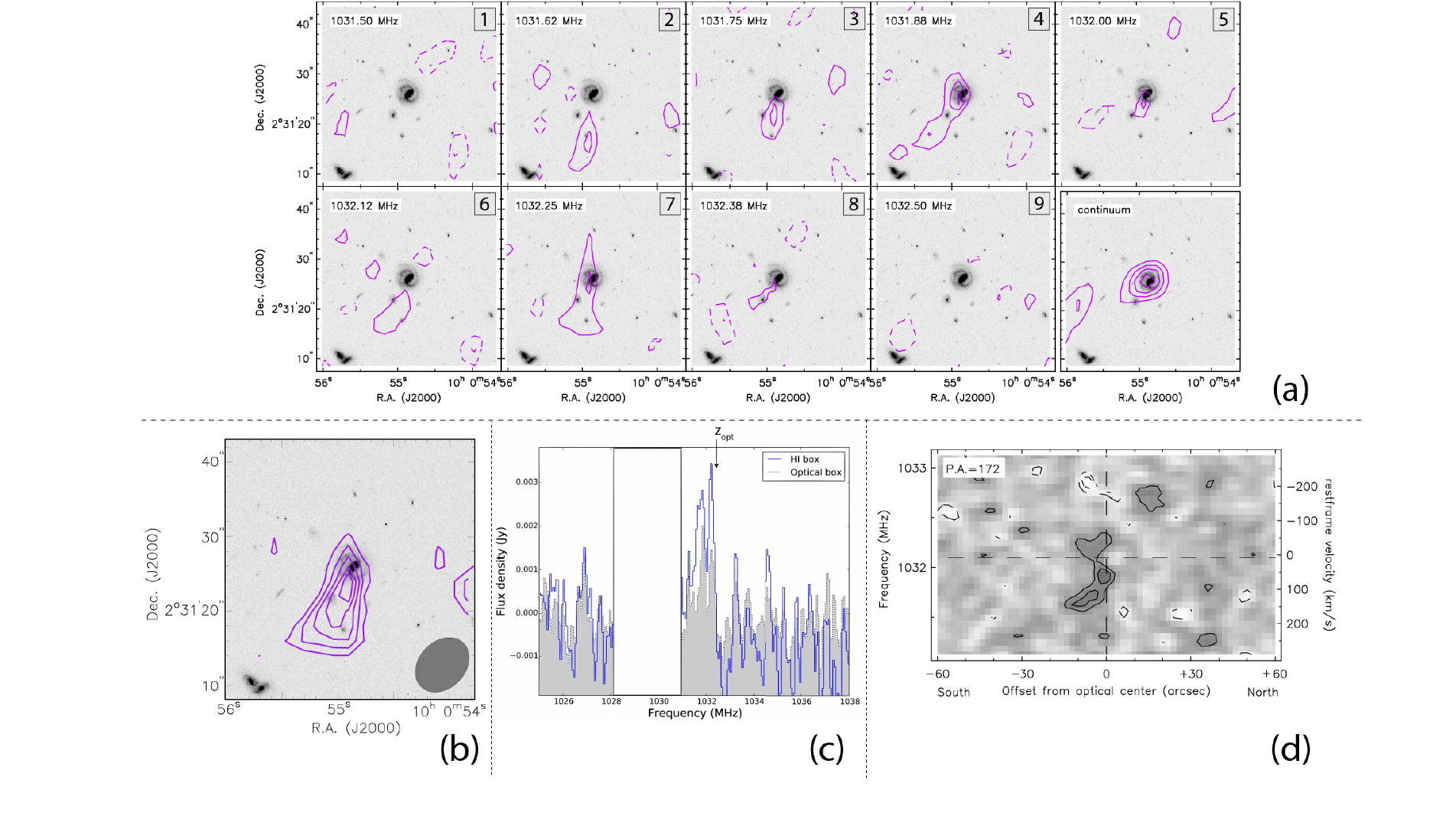}
\caption{\HI detection of J100054. (a) Channel maps showing \HI contours drawn in levels of  $\pm 2\sigma, \pm~3\sigma$, $+4\sigma$ (negative contours are dashed). The last panel shows the continuum emission for this galaxy.  (b) Total \HI distribution map. The ellipse shows the size of the synthesized beam ($8\farcs2\times6\farcs3$), and the contours correspond to $(0.4, 0.9, 1.3, 1.7, 2.1)  \times 10^{20}$ atoms cm$^{-2}$. The background optical data is an HST I-band image from COSMOS \citep{HST1, HST2}. (c) \HI spectra integrated over the optical disk of the galaxy (gray) and over most of the \HI emission (blue). The blanked region corresponds to frequencies affected by RFI and the arrow shows the frequency corresponding to the optical redshift.
(d) PV diagram made through the optical center at P.A.$=172\degree$ with contours drawn in levels of  $\pm 2\sigma, \pm~3\sigma$.  The vertical line corresponds to the optical position of the galaxy and the horizontal line to a frequency of 1032.1 MHz.}  
\label{fig:HIfig}
\end{center}
\end{figure*}

We detect \HI in the galaxy J100054, which has a predicted $M_{\rm HI}=2.8\times10^{10}~M_\odot$. The \HI channel maps (Figure \ref{fig:HIfig}a) show that the emission is mostly at the $2-3\sigma$ level (with a $4\sigma$ peak at the position of the galaxy). Although the emission is weak, we trust the detection since the signal is at the location and velocity of the galaxy, and  appears in several consecutive independent channels.  The emission is seen in 7 panels making it 875 kHz wide, which translates to a velocity width of $246 \pm 36$ \kms after correcting for the channel width. We calculate $z_{\rm HI}=0.3764\pm0.0002$ assuming that the center of the emission is in panel 5.  We have high confidence in the emission seen in panels 2-5 since it is at the $3\sigma$ level, with a smooth velocity gradient, and there are no negative contours of high significance. The emission in panels 6-8 is weaker and its velocity does not follow a clear pattern.   We generate the total \HI distribution map (Figure 1b) by adding the emission seen in the 7 channels after smoothing and applying a $1.25\sigma$ cutoff. We also make a moment map excluding the emission seen in panels 6-8, and the morphology is almost identical.  We calculate $M_{\rm HI}$ for both maps, $M_{\rm HI}=(2.9\pm1.0)\times10^{10}~M_\odot$ for the one presented here and $M_{\rm HI}=2\times10^{10}~M_\odot$ from the map only including panels 2-5, showing that the \HI in panels 6-8 does not contribute much.  As seen in Figure 1b, the \HI is asymmetric and very extended, encompassing some potentially  nearby companions.  Since the low-level emission in the southern extension is rather sensitive to the noise, the morphology is uncertain.  We also include two integrated \HI spectra in Figure 1c, one that integrates over the optical disk of the galaxy, and another one that is centered on the \HI emission.  We calculate a S/N=7 from the \HI spectrum by adding up the signal in the range 1031.5-1032.5 MHz, and dividing by the rms calculated in the range 1033-1038 MHz (taking into account the number of channels).  Lastly, Figure 1d is a position velocity (PV) diagram that further shows the significance of the detection.

The last panel of Figure 1a shows the continuum emission from the line-free channels.  In addition,  we have continuum data from the commensal survey CHILES Continuum Polarization (CHILES Con Pol)\footnote{\url{http://www.chilesconpol.com}}. The source has a total flux density of $240 \pm 10$ $\micro$Jy at 1.45 GHz, which is consistent with our measurement and the COSMOS-VLA data \citep{Schinnerer07}. We measure a spectral index of  $-0.8 \pm 0.15$, which is the typical value for star-forming galaxies \citep[e.g.][]{Magnelli}.

\subsubsection{CO Observations with the LMT}
As a follow-up to our \HI detection, we observed J100054 using the Redshift Search Receiver (RSR) at the LMT \citep{Erickson07} on April 8 and May 21, 2015 for 1 hr each, under excellent weather conditions ($T_{sys}\equiv 100$ K). The spectrum covering a frequency range $73-111$ GHz is obtained with a spectral resolution of 31.25 MHz ($\sim$110 km s$^{-1}$ at $z=0.376$) to search for possible CO(1--0) emission. The effective beam size is $25\arcsec$ at 84 GHz, and telescope pointing is checked on the nearby QSO J0909+013 before each observing session.  When averaged together with the $1/\sigma^2$ weight, the final spectrum has an rms noise of $\sigma=1.2$ mJy.  

\begin{figure}
\begin{center}
\includegraphics[trim=10mm 28mm 0mm 0mm,clip,scale=0.48]{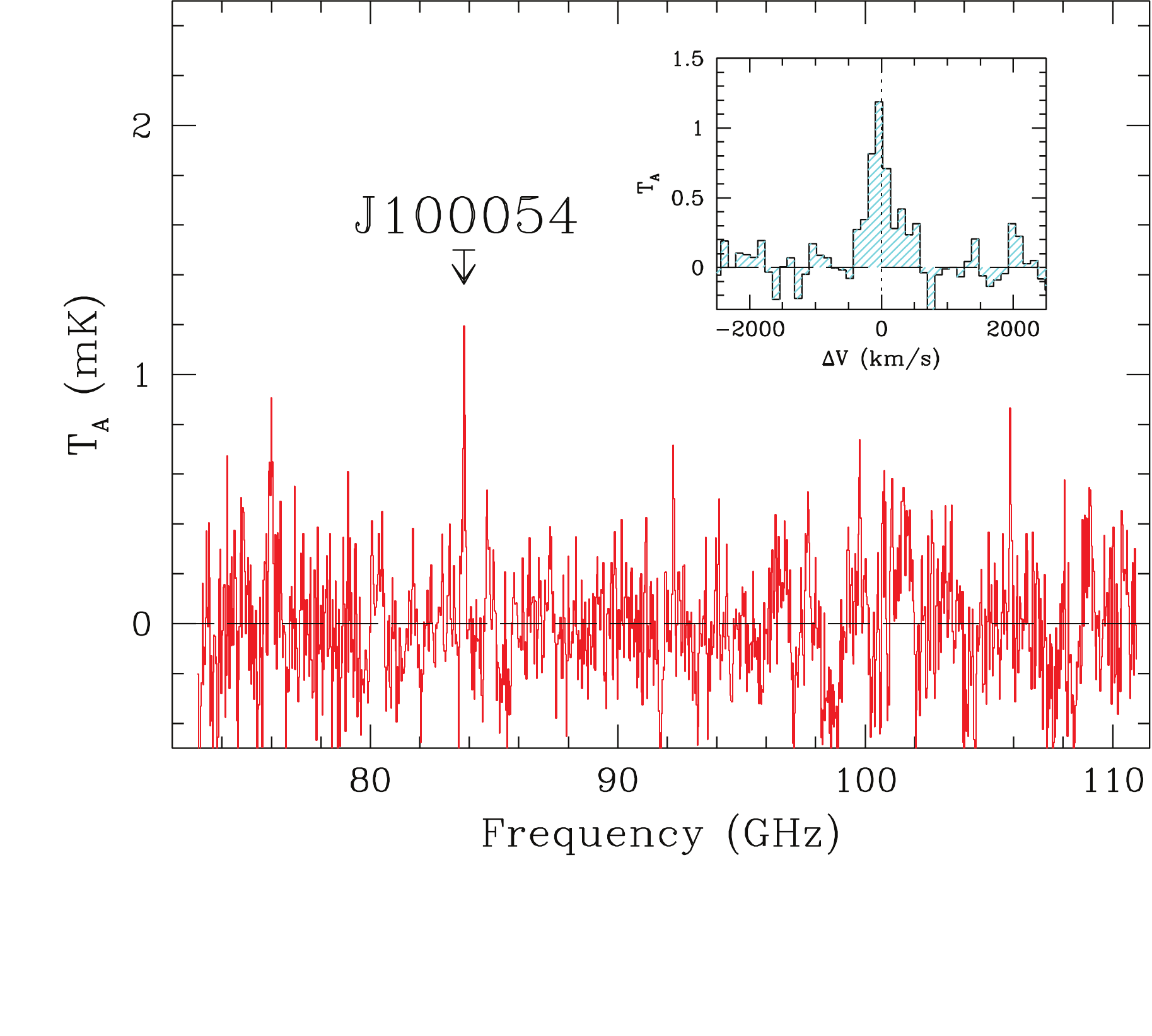}
\caption{CO spectrum for J100054 obtained using the RSR on the LMT. The larger panel shows the full frequency range, and the inset shows a zoom-in on the CO(1--0) detection at 83.7727 GHz ($z=0.376$).} 
\label{fig:highzCO}
\end{center}
\end{figure}

A single bright line is detected with a S/N = 9.7 centered at 83.7816 GHz (Fig.~\ref{fig:highzCO}), and is interpreted as that of the CO(1--0) transition at $z=0.3759\pm0.0002$.  This line is fully resolved, as shown by the zoom-in spectrum shown in the inset, and the derived FWHM is $413\pm62$ km s$^{-1}$ after correcting for the instrumental resolution.  The narrow CO line peak (FWHM is $317\pm56$ km s$^{-1}$), centered on the optical redshift $z=0.376$, sits atop a broad base, which accounts for the observed broad line width. The measured CO line integral of $3.12\pm0.32$ Jy km s$^{-1}$ translates to $L_{\rm CO}=(2.3\pm0.2) \times 10^{10}$ K km s$^{-1}$ pc$^2$. The H$_2$ mass depends on the adopted conversion factor, which ranges from $\alpha=0.8 \,M_\odot$/(K \kms\ pc$^2$) for interacting galaxies \citep{Downes98} to the Galactic conversion factor of  $\alpha=4.3\,M_\odot$/(K \kms\ pc$^2$) \citep{Bolatto13}. This results in a range of possible values of  $M_{\rm H_2}=(1.8-9.9)\times10^{10}~M_\odot$. 
    
\begin{figure*}
\begin{center}
\includegraphics[trim=0mm 0mm 22mm 0mm,clip, scale=0.9]{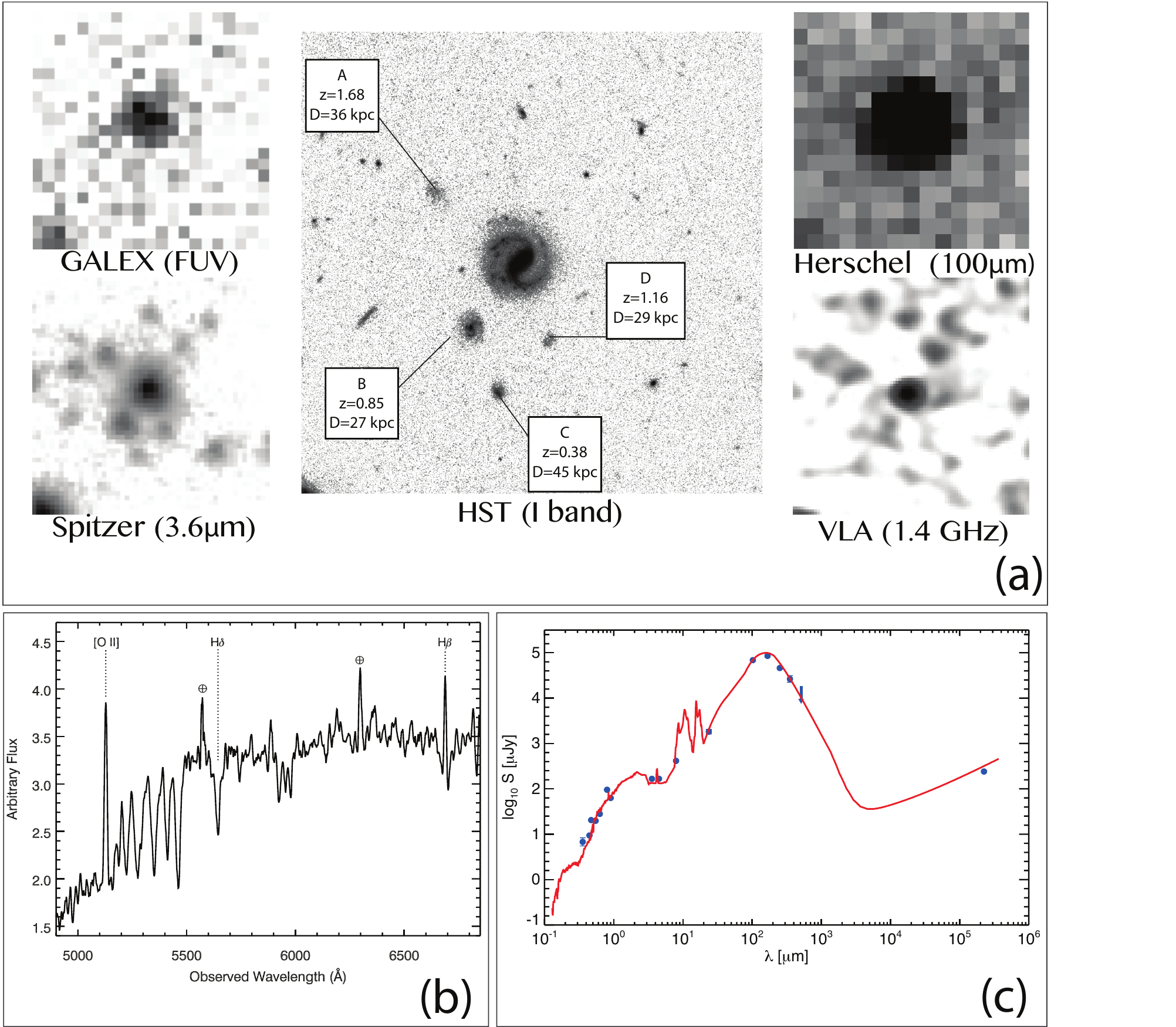}
\caption{Multiwavelength data for J100054. (a) Images of J100054 detected with different telescopes: GALEX, Spitzer, HST, Herschel and the VLA (all are 30$\arcsecond\times 30\arcsecond$). The central image is labeled with possible companions, their photometric redshift, and projected distance from the main galaxy (assuming they are at the same redshift). (b) Optical spectrum that confirms the redshift of the galaxy, the different lines are labeled and the earth symbols correspond to prominent sky lines. (c) SED fit using the available photometry. } 
\label{fig:optical}
\end{center}
\end{figure*}

\subsection{Additional data}

\subsubsection{COSMOS multiwavelength data}
As J100054 is located within the COSMOS field, extensive multiwavelength data are available, and it is detected in ultraviolet (UV), optical, IR, and by the VLA-COSMOS continuum survey. A deep HST I-band image of J100054 (Figure 3a) shows that it is a massive regular barred spiral with clear spiral arms and a prominent bulge. In addition, it shows a number of fainter galaxies suggesting that this may be a small group. The information for the possible companions is scarce, as the four are very faint with $r>23$ and only have measured photometric redshifts. Only one of the companions has a photometric redshift similar to J100054, but spectroscopic data are necessary to confirm this and determine whether the other ones are indeed background objects.

\subsubsection{Optical spectroscopic confirmation} 
We obtained an additional long-slit spectrum with the SOAR
4.1m telescope and the Goodman High Throughput Spectrograph on 2015 April 17 to confirm the redshift of J100054. A slit of 1.68$^{\prime\prime}$ width was oriented north-south on the galaxy, and the 400 l/mm grating yielded spectral coverage 3500--7500\AA~ at a resolution of $R=495$. The 1 hour exposure confirms a galaxy at $z=0.3758 \pm 0.0001$ with strong [OII] 3727\AA~ emission and Balmer absorption, indicating a post-starburst stellar population. The spectrum shows no signs of AGN activity, as seen in Figure 3b.

\subsubsection{Spectral Energy Distribution (SED) Fitting}
The optical and IR counterparts are found exploiting the likelihood ratio technique \citep{Sutherland92} with the optical photometric data \citep{Capak07}, Spitzer Space Telescope MIPS 24 $\mu$m \citep{LeFloch09} and Herschel Space Observatory \citep{Lutz11,Oliver11}. The best-fit SED is obtained through the $Le$ $Phare$ \citep{Arnouts99, Ilbert06} with the M82 galaxy template for the optical-mid IR regime and the Chary-Elbaz galaxy template for the far-IR regime at the fixed redshift of $z=0.37$. The best-fit SED yields the total  $L_{\rm IR}$($8-1000$ $\micron$)$=(5.7\pm0.4) \times 10^{11} L_{\odot}$ and the $L_{\rm FIR}$($40-120$ $\micron$)$=(3.0\pm0.2) \times 10^{11}$ $L_{\odot}$, confirming that this object is a LIRG.  

\subsubsection{Stellar Mass and SFR Estimate}
We compute $M_*=(8.7\pm0.1) \times 10^{10}$ $M_{\odot}$ using the Spitzer IRAC data \citep{Eskew12}. Following  \citet{Murphy11},  we calculate the SFR using the FUV, IR, and radio data: SFR$_{\rm FUV}=$ ($0.7\pm0.1$) $M_{\odot}$ yr$^{-1}$, SFR$_{\rm IR}=$ ($85\pm6$) $M_{\odot}$ yr$^{-1}$, and SFR$_{\rm 1.4GHz}=$ ($72\pm3$) $M_{\odot}$ yr$^{-1}$ (errors come from the photometric uncertainties). The SFR$_{\rm FUV}$ is negligible, indicating that most of the star-formation is dust-obscured.  We adopt SFR$_{\rm IR}$ for the rest of the paper.  

\section{Discussion}
\subsection{Unusual spiral galaxy}
While the optical image shows J100054 to be a large normal spiral, the cold gas properties depict a more complex picture.  The \HI mass agrees with the predicted one, but its distribution and kinematics are unexpected.
The \HI suggests the possibility of a  gravitational interaction, as the \HI
looks very disturbed. Many studies have reported SFR enhancement in interacting pairs. In particular, in the case of a minor merger, the more massive companion experiences a stronger SFR enhancement than the less massive one \citep[e.g.,][]{Scudder12,Davies15}.  Spectroscopic data are necessary to determine whether the nearby galaxies are at the same redshift.

In addition, the galaxy is extremely H$_2$ rich, with a value that is even unexpected for local (U)LIRGs. The $L_{\rm CO}$ for this system is approximately 6 times higher than objects with similar $L_{\rm IR}$  in a sample of local (U)LIRGs (G. Privon, in prep). Here we present a range of possible values for $M_{\rm H_2}$ (depending on $\alpha$) but we note that $\alpha=0.8$ is still a matter of debate even for extreme (U)LIRGs such as Arp 220 \citep[][]{Scoville15}.
The CO spectrum suggests there might be two components, a broad one and a narrow one.  The FWHM of the narrow component is consistent with the \HI width within the uncertainties. The broad component could be emission from neighboring galaxies included in the large beam. We note, however, that the galaxies are fairly small and we do not expect significant contribution. The more likely explanation is an outflow such as those seen for local (U)LIRGs \citep[e.g.][]{Chung11}. 

\subsection{Comparison with other samples}
\begin{figure*}
\begin{center}
$\begin{array}{cc}
\includegraphics[trim=8mm 6mm 17mm 18mm,clip,scale=0.38]{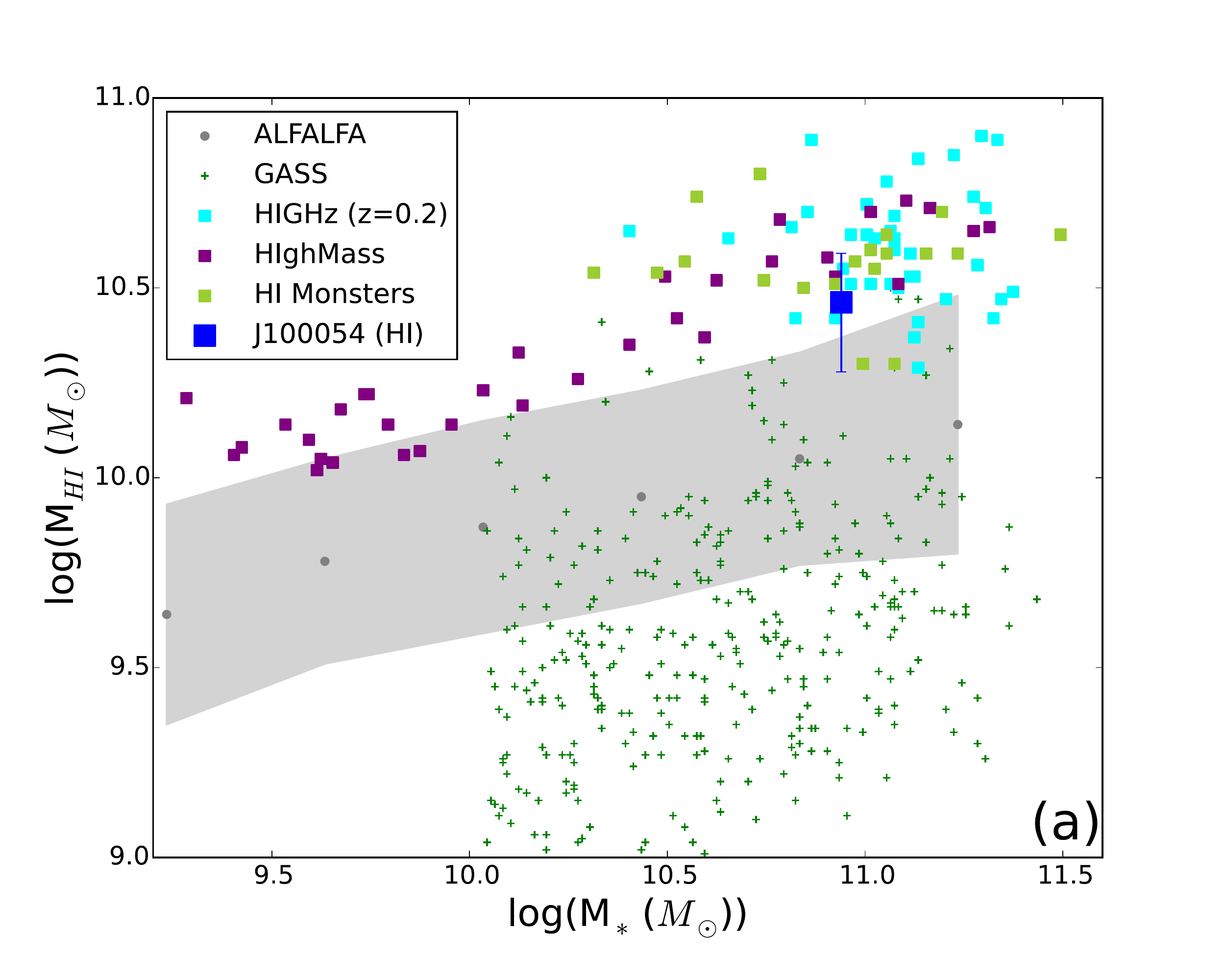}&
\includegraphics[trim=8mm 6mm 17mm 18mm,clip,scale=0.38]{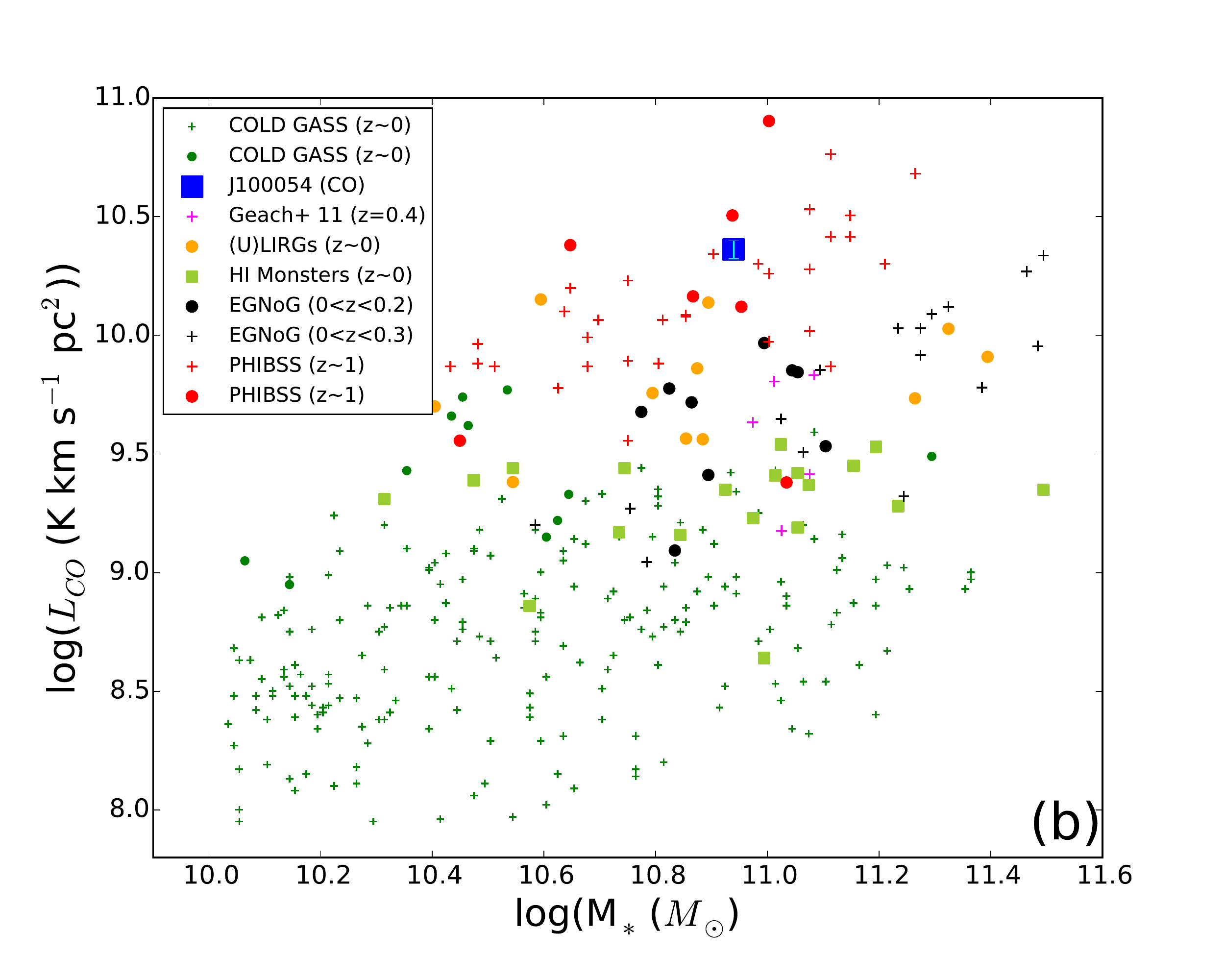}\\
\includegraphics[trim=8mm 6mm 17mm 18mm,clip,scale=0.38]{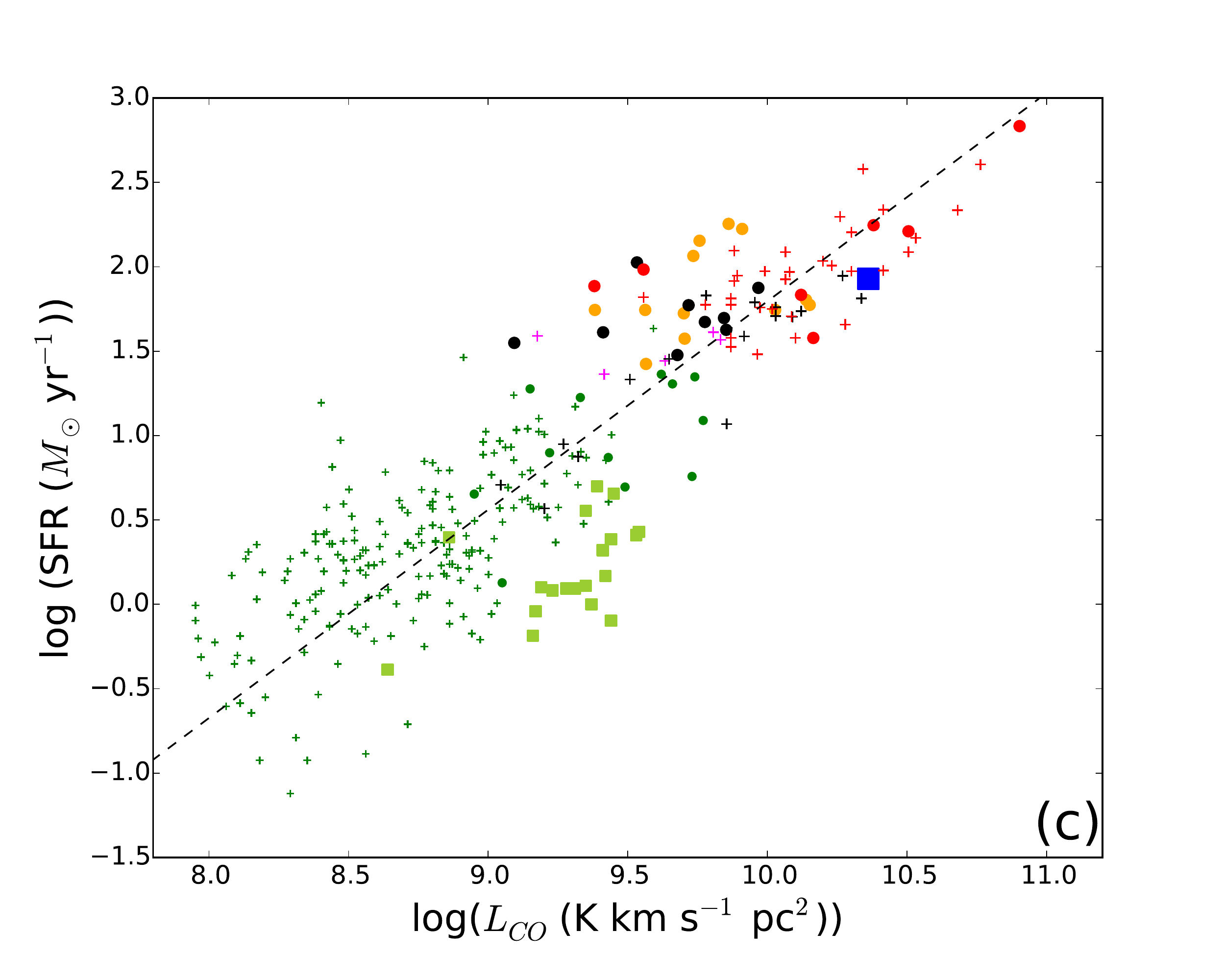}&
\includegraphics[trim=8mm 6mm 17mm 18mm,clip,scale=0.38]{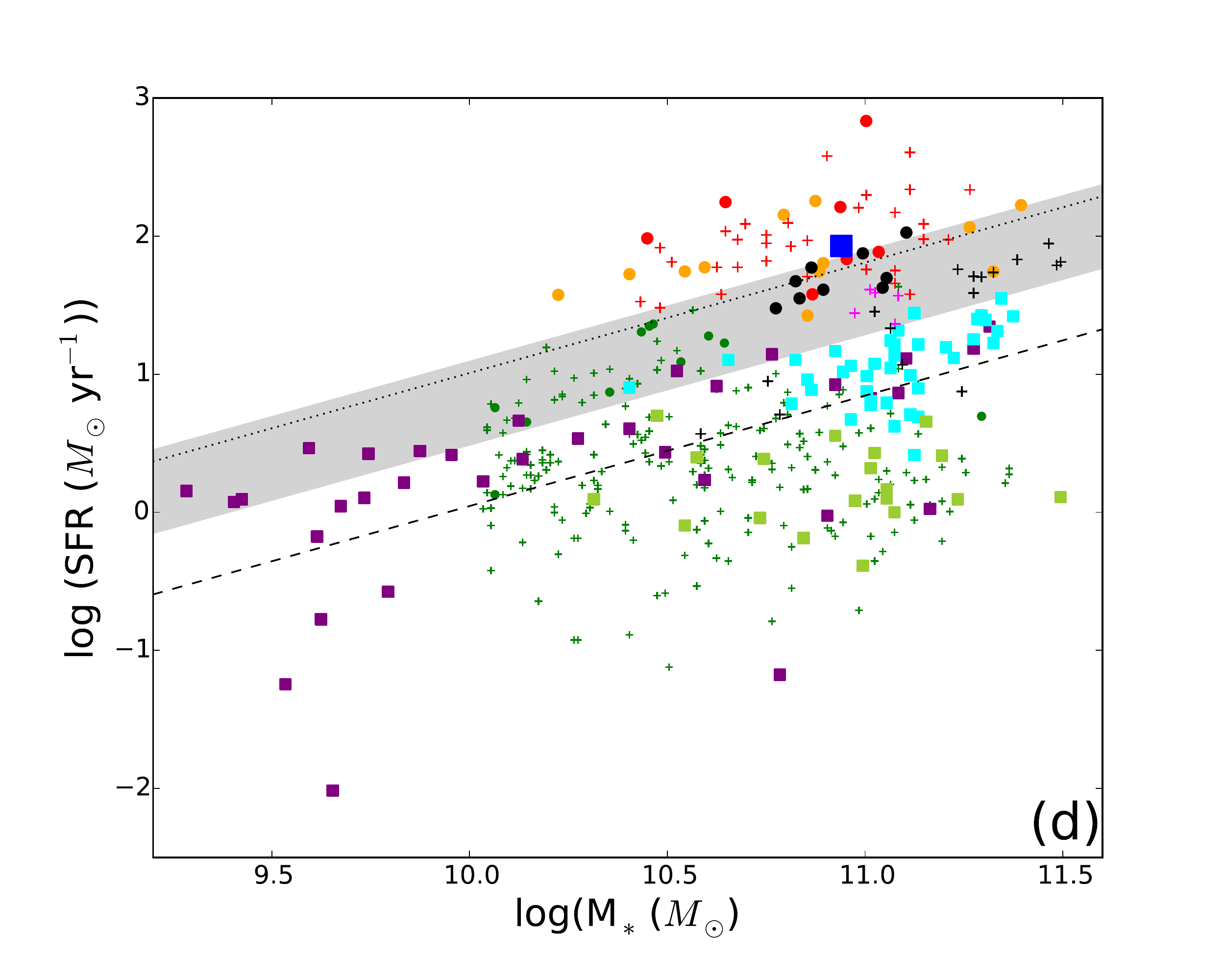}
\end{array}
$
\caption{Properties of J100054 compared to other surveys. The different colors correspond to: blue (results presented here), gray \citep[ALFALFA; median and 1$\sigma$ values for each bin from][]{Maddox15}, purple \citep[HIghMass selected from ALFALFA;][]{Huang14}, yellow-green \citep[\HI monsters;][]{Lee14}, cyan \citep[HIGHz;][]{Catinella15}, green \citep[GASS and COLD GASS;][]{Catinella12,Saintonge11}, magenta \citep{Geach11}, orange (CO from \citet{Gao04}, $M_*$ and SFR from \citet{Vivian}), black \citep[EGNoG;][]{Bauer13}, and red \citep[PHIBSS;][]{Tacconi13}. The filled circles represent mergers, starbursts and/or ULIRGs, the plus signs correspond to normal star-forming galaxies, and the squares are unclassified. The different redshifts are noted in the legends. All of the numbers shown here are calculated using the same cosmology and Kroupa IMF. The $L_{\rm CO}$ correspond to CO(1-0) observations, except for PHIBSS that observed CO(3-2) but we convert to CO(1-0) assuming CO(1-0)/CO(3-2)=2.  (a) $M_{\rm \HI}$  vs $M_{*}$.  (b) $L_{ \rm CO}$ vs $M_{*}$. (c) SFR vs $L_{\rm CO}$. The dashed line corresponds to the correlation between $L_{\rm CO}$ and $L_{\rm IR}$ for normal star-forming galaxies \citep{Sargent14}.  (d) SFR vs $M_{*}$ compared to the SFR-$M_*$ relation at different redshifts from \citet{Bauer13}. The shaded region corresponds to $z=0.376$, with the lower bound showing the SFR-$M_*$ relation at that $z$ and the upper bound indicating the starburst threshold. The relations at $z=0$ (dashed line) and at $z=1$ (dotted line) are  included for reference. }

\label{fig:comparison}
\end{center}
\end{figure*}

 Figure 4 explores whether the gas properties are unusual for a galaxy with that $M_*$ and SFR when compared to the properties of galaxies
found in other surveys (see caption for details and references). Figure \ref{fig:comparison}a compares the \HI mass and $M_*$ of J100054 to systems
found in other surveys.  As seen, J100054 is rich in atomic gas when compared to ALFALFA and GASS, and in fact has as much \HI as the most \HI-rich galaxies at $z\sim0-0.2$. Figure \ref{fig:comparison}b shows $L_{\rm CO}$ (instead of H$_2$ mass to avoid issues with $\alpha$) as a function of $M_*$. The plot shows that most galaxies have lower $L_{\rm CO}$, only  those at $z\sim1$ have comparable values to J100054. The third plot shows SFR as a function of $L_{CO}$ (Figure \ref{fig:comparison}c).  This is usually plotted as $L_{\rm IR}$ vs $L_{\rm CO}$, but here we choose SFR since some of the samples do not have published $L_{\rm IR}$.  From this plot, we can infer that J100054 has a relatively low star formation efficiency since it lies below the line that marks the SFR-$M_*$ relation (commonly known as the ``main sequence").  Lastly, we look at the SFR for different $M_*$ (Figure \ref{fig:comparison}d).  J100054 is just above the starburst threshold, which is defined as four times the SFR-$M_*$ relation \citep[][]{Rodi11}. In summary, these four plots show that J100054 is gas-rich both in molecular and atomic gas, but its SFR is somewhat low given its $L_{\rm CO}$. 

\vspace{0.2in}

\subsection{ISM studies at intermediate redshift}
Lastly, we emphasize this is the first time we observe both the CO and \HI from a galaxy beyond the local Universe ($z\sim0.2$). In Figure 5 we show how the gas fraction  and $M_{H_2}$/$M_\HI$ ratio vary with redshift for three surveys: COLD GASS, BUDHIES \citep{Verheijen07,Cybulski15}, and CHILES. These limited data suggest that if we assume $\alpha=4.3$, both the gas fraction and
ratio of H$_2$/\HI go up with increasing redshift and that J100054 is above the median
seen in surveys at $z=0$ and $z=0.2$. If we adopt $\alpha=0.8$, the gas fraction is still high when compared to other surveys, but the mass ratio is comparable to galaxies at $z=0$.   We can not draw strong conclusions with only one data point but we will soon be able to start populating these plots with upcoming CHILES results and follow-up observations with the LMT.  This will be further complemented by the upcoming surveys that will be conducted with ALMA, and the SKA and its precursors in the next decade. 

In addition to probing the general galaxy population, we can also start to probe the evolution of (U)LIRGs.  We know these systems correspond to mergers in the local Universe, but there is disagreement on the merger contribution at higher redshifts \citep[e.g.,][]{Kartaltepe10}.  Future \HI and H$_2$ data will allow us to understand these systems better and probe their ISM across cosmic time.   

\begin{figure*}
\begin{center}
$\begin{array}{cc}
\includegraphics[trim=8mm 6mm 17mm 18mm,clip,scale=0.38]{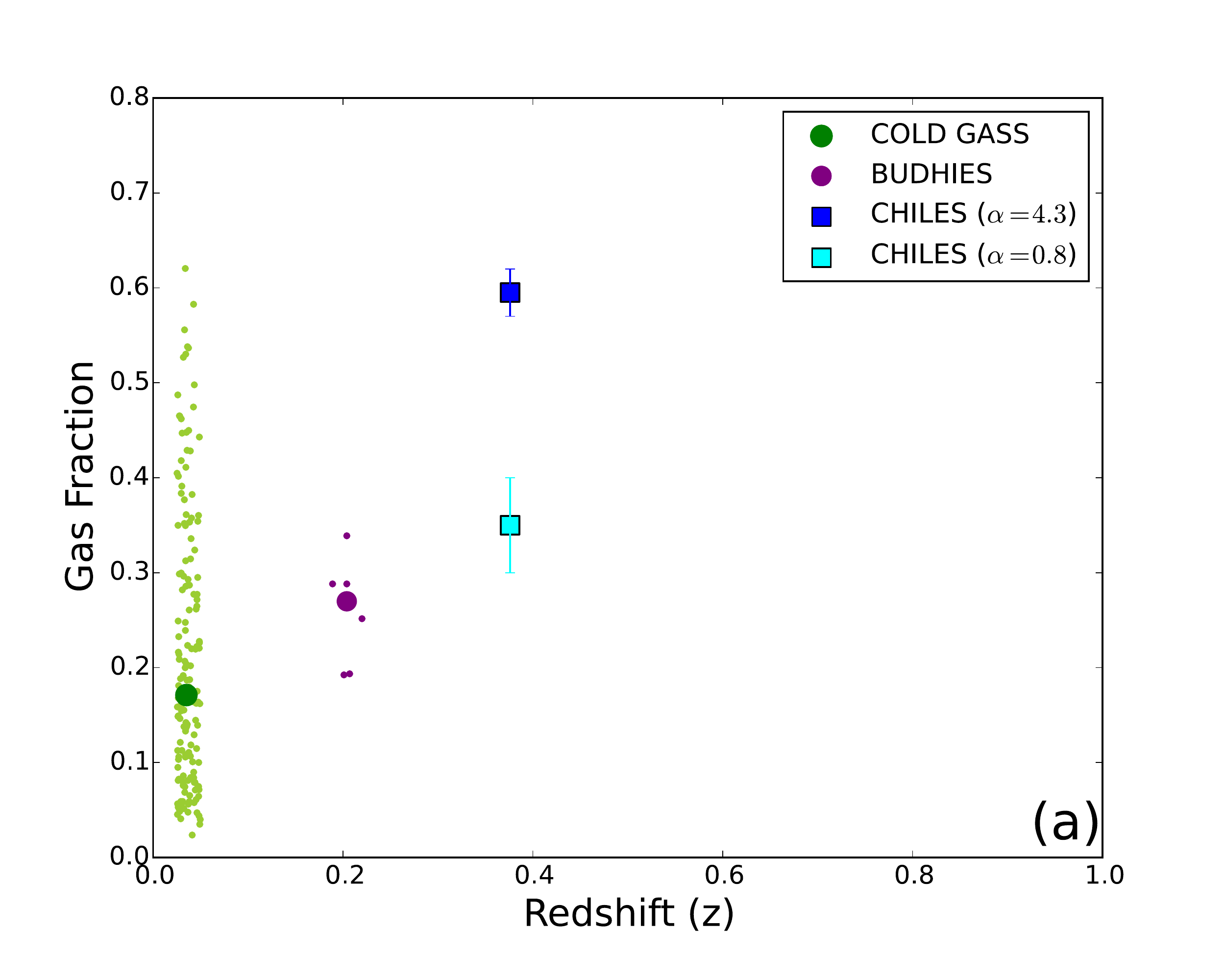}&
\includegraphics[trim=8mm 6mm 17mm 18mm,clip,scale=0.38]{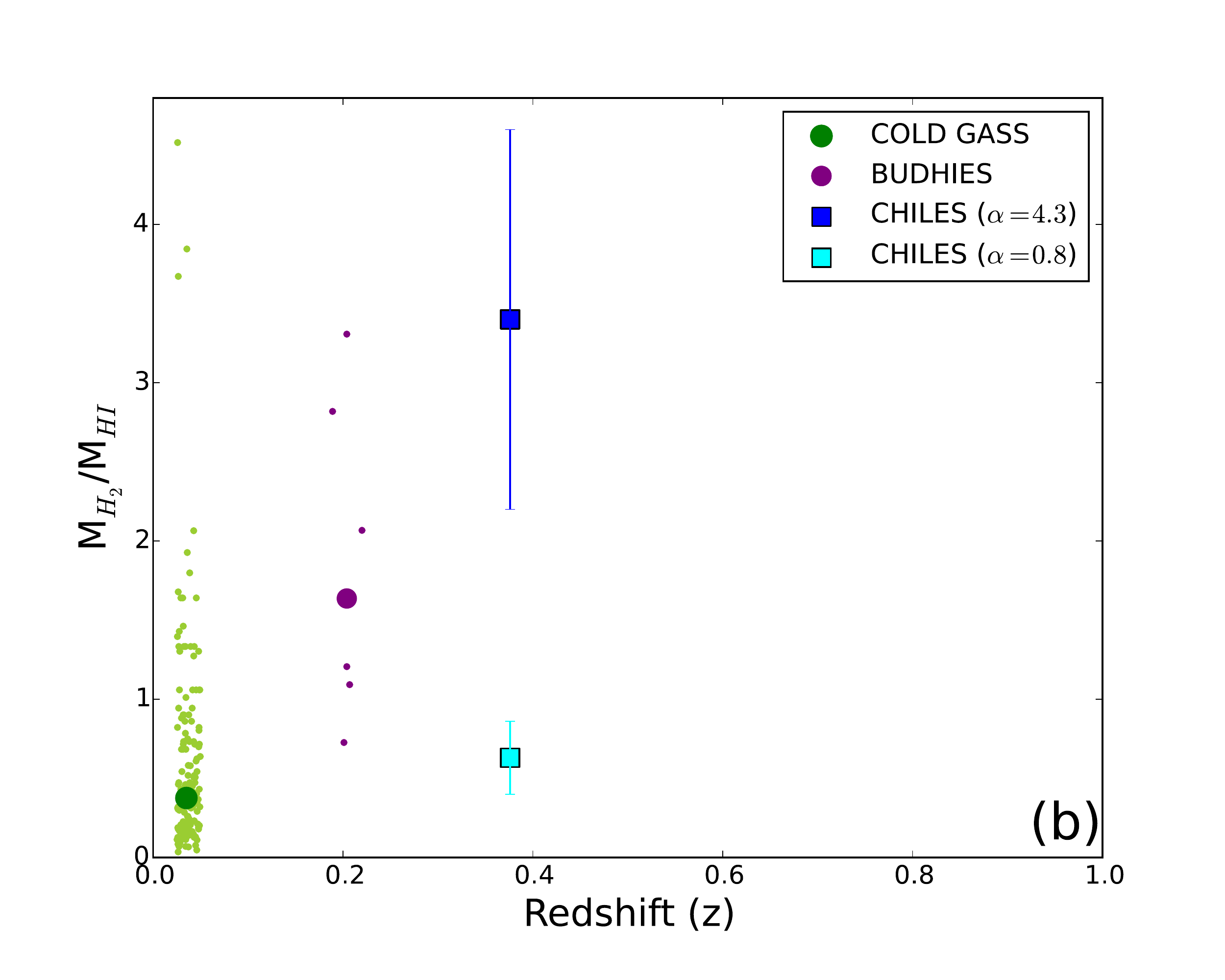}
\end{array}
$
\caption{ISM properties at different redshifts for three surveys: COLD GASS, BUDHIES and CHILES. We include all of the galaxies with detected \HI and $H_2$ for the first two surveys, with the bigger symbol showing their median value.  (a) Gas fraction (($M_{\rm \HI}$+$M_{\rm H_2}$)/($M_{\rm \HI}$+$M_{\rm H_2}$+$M_*$)) vs. redshift. (b) H$_2$/\HI ratio vs. redshift.}
\end{center}
\end{figure*}

\section{Summary}
We presented the first comprehensive study of the gas content (\HI and H$_2$) of a galaxy at intermediate redshifts. We summarize our main findings below:
\begin{itemize}
\item We detected the highest redshift \HI in emission to date from a very gas-rich system ($M_{HI}=(2.9\pm1.0)\times10^{10}~M_\odot$).  Its \HI mass is similar to the most gas-rich galaxies locally known  and consistent with what is expected from its stellar properties.
\item J100054 is also gas-rich in H$_2$, with a  mass range of $(1.8-9.9)\times10^{10}~M_\odot$. The CO luminosity is higher than what is expected for a galaxy with that stellar mass ($8.7 \times 10^{10}$ $M_{\odot}$) and SFR (85 $M_{\odot}$ yr$^{-1}$).
\item In comparison to other samples, the CO properties suggests J100054 is similar to star-forming galaxies at $z\sim1$.
\end{itemize}

\acknowledgments
We thank the referee for constructive feedback that helped us improve the paper.  We acknowledge useful discussions with Andrew Baker and George Privon. CHILES is supported by NSF grants AST-1413102, AST-1412578,  AST-1412843, AST-1413099 and AST-1412503. XF is supported by an NSF-AAPF under award AST-1501342. LC and ET acknowledge support from NSF grant AST-1412549. KH has been supported by the ERC under the EU's Seventh Framework Programme (FP/2007-2013)/ERC Grant Agreement nr.~291531. KK acknowledges grants KR 4598/1-2 and SCHI 536/8-2 from the DFG Priority Program 1573. YJ acknowledges the Marie Curie Actions of the European Commission (FP7-COF). 
This work would not have been possible without the long-term financial support from the Mexican Science and Technology Funding Agency, CONACYT during the construction and early operational phase of the Large Millimeter Telescope Alfonso Serrano, as well as support from the the US NSF via the URO program, the INAOE and UMASS-Amherst. 
Based on observations obtained at the SOAR telescope, which is a joint project of the MCTI da Rep\'{u}blica Federativa do Brasil, NOAO, UNC-Chapel Hill, and MSU.

\end{document}